\documentclass[aps, preprint, eqsecnum, showpacs, epsfig]{revtex4}
\usepackage{graphics}
\usepackage{graphicx}
\usepackage{epsfig}
\def \PY {Percus-Yevick}

\def \be {\begin{equation}}
\def \ee {\end{equation}}
\def \ba {\begin{eqnarray}}
\def \ea {\end{eqnarray}}
\def \bm {\begin{displaymath}}
\def \em {\end{displaymath}}
\def \br {{\bf r}}
\def \bom {{\bf \Omega}}
\def \bn {{\bf 1}}
\def \bp {{\bf 2}}
\def \bs {{\bf 3}}

\begin{document}
\title{Freezing transitions in fluids of long elongated molecules}
\author{Pankaj Mishra, Jokhan Ram and Yashwant Singh}
\affiliation{Department of Physics, Banaras Hindu University, 
Varanasi-221 005,
India}
\date{\today}
\begin{abstract}
We have used the density-functional theory to locate the freezing transitions and 
calculate the values of freezing parameters for a system of long elongated 
molecules which interact via the Gay-Berne pair potential. The pair correlation
functions of isotropic phase which enter in the theory as input informations are 
found from the Percus-Yevick integral equation theory. At low temperatures 
the fluid freezes directly into the smectic $A$ phase on increasing the 
density. The nematic phase is found to stabilize in between the isotropic and 
smectic $A$ phases only at high temperatures and high densities. These features of 
the phase diagram are in good agreement with the computer simulation results.
\end{abstract}
\maketitle
\section{\bf Introduction}
A system consisting of anisotropic molecules is known to exhibit liquid crystalline 
phases in between the isotropic liquid and the crystalline solid. The liquid crystalline 
phases that commonly occur in a system of long elongated molecules are nematic and smectic
(Sm $A$) phases \cite{1}. In the nematic phase the full translational symmetry of the 
isotropic fluid phase (denoted as $R^3$) is maintained but the rotational 
symmetry $O(3)$ or $SO(3)$ (depending upon the presence or absence of the 
center of symmetry) is broken. In the simplest form of the axially symmetric 
molecules the group $SO(3) $ [or $O(3)$] is replaced  by one of the uniaxial symmetry 
$D_{\infty h}$ or $D_\infty$. The phase possessing the $R^3 \wedge D_{\infty h} $ 
(denoting the semi-direct product of the translational group $R^3 $ and the rotational group 
$D_{\infty h}$) symmetry is known as uniaxial nematic phase. 

The smectic liquid crystals, in general, have a stratified structure with the long 
axis of molecules parallel to each other in layers. This situation corresponds to partial breakdown 
of translational invariance in addition to the breaking of the rotational 
invariance. Since a variety of molecular arrangements are possible within each layer, a 
number of smectic phases are possible. The simplest among them is the Sm $A$ phase.
In this phase the center of mass of molecules in a layer are distributed as in a two dimensional
fluid but the molecular axes are on the average along a direction normal
to the smectic layer (i.e. the director ${\bf\hat n}$ is normal to the smectic layer).
The symmetry of the Sm $A$ phase is $(R^2\times Z)\wedge D_{\infty h}$ 
where $R^2$ corresponds to a two dimensional liquid structure and Z for a one-dimensional 
periodic structure. The other non-chiral and non-tilted smectic phase seen in a system 
of long elongated molecules is smectic $B_h$(Sm $ B_h$) phase. In each smectic layer of the 
Sm $B_h$ phase, the director is parallel to the layer normal as in the Sm $A$ phase, and 
there is short-range positional but long range bond-orientational hexagonal orders in smectic
plane. The azimuthally symmetrical x-ray ring of Sm $A$ is replaced by a six-fold 
modulated diffused pattern \cite{2}. The phase is thus characterized by a $D_{6h}$ point group 
symmetry and is uniaxial like Sm $A$.

All these phases including that of the isotropic liquid and the crystalline solids 
are characterized by the average position and orientation of molecules and by
the intermolecular spatial and orientational correlations. The factor responsible for 
the existence of these distinguishing features is the anisotropy in both the shape of molecules 
and the attractive forces between them. The relationship between the intermolecular 
interactions and the relative stability of these phases is very intriguing and not 
yet fully understood. For a real system one faces the problem of knowing accurate 
intermolecular interaction as a function of intermolecular separation 
and orientations. This is because the mesogenic molecules 
are so complex that none of the methods used to calculate interactions between 
molecules can be applied without drastic approximations. Consequently, one 
is forced to use the phenomenological descriptions, either as a straightforward 
model unrelated to any particular physical system , or as a basis for describing by means of adjustable
parameters between two molecules. Since our primary interest here is to relate 
the phases formed and their properties to the essential molecular factor responsible for the 
existence of liquid crystals and not to calculate the properties of any real system,
the use of the phenomenological potential is justified. One such phenomenological model 
which has attracted lot of attention in computer simulations is one proposed by Gay and 
Berne \cite{3}. 

In the Gay-Berne (GB) pair potential model, the molecules are viewed as rigid units 
with axial symmetry. Each individual molecule $i$ is represented by a center-of-mass
position ${\bf r_i}$ and an orientation unit vector ${\bf \hat e_i}$ which is in the 
direction of the main symmetry axis of the molecule. The GB interaction energy between a pair of 
molecules (i, j) is given by 
\begin{equation}
u({\bf r_{ij}},{\hat {\bf e_i}}, {\hat {\bf e_j}}) =
4 \epsilon({\hat {\bf e_i}}, {\hat {\bf e_j}})
(R^{-12}-R^{-6})
\end{equation}
where
\begin{equation}
R = \frac{r_{ij}-\sigma({\hat{\bf e_i}}, {\hat{\bf e_j}}, {\hat\br_{ij} })+\sigma_0}{\sigma_0}
\end{equation}
Here $\sigma_0$ is a constant defining the molecular diameter, $r_{ij}$ is the 
distance between the center of mass of molecules $i$ and $j$ and 
$\hat\br_{ij}={ \br_{ij}}/{|\br_{ij}|}$ is a unit vector along the center-center 
vector $\br_{ij}= \br_i-\br_j$. $\sigma({\hat{\bf e_i}}, {\hat{\bf e_j}}, {\hat\br_{ij} })$ 
is the distance (for given molecular orientation) at which the intermolecular 
potential vanishes and is given by 
\begin{widetext}
\ba
\sigma({\hat{\bf e_i}}, {\hat{\bf e_j}}, \hat\br_{ij})& =&
\sigma_0 \left[ 1 - \chi\left(\frac{({\hat{\bf e_i}}.{\hat\br_{ij}})^2
+ ({\hat{\bf e_j}}.{\hat\br_{ij}})^2 - 2\chi({\hat{\bf e_i}}.
{\hat\br_{ij}})({\hat{\bf e_j}}.{\hat\br_{ij}})
({\hat{\bf e_i}}.{\hat{\bf e_j}})}
{1 - \chi^2({\hat{\bf e_i}}.{\hat{\bf e_j}})^2}\right)\right]^{-\frac{1}{2}}
\ea
\end{widetext}
The parameter $\chi$ is a function of the ratio $x_0(\equiv \sigma_e/\sigma_s)$
which is defined in terms of the contact distances when the particles are end-to-end 
$(e)$ and side-by-side $(s)$,
\be
\chi=\frac{x_{0}^{2}-1}{x_{0}^{2}+1}
\ee
The orientational dependence of the potential well depth is given by a product of 
two functions 
\begin{equation}
 \epsilon({\hat {\bf e_i}}, {\hat {\bf e_j}},\hat\br_{ij})=
\epsilon_0 \epsilon^{\nu}({\hat {\bf e_i}}, {\hat {\bf e_j}})
\epsilon'^{\mu}({\hat{ \bf e_i}}, {\hat {\bf e_j}},\hat\br_{ij})
\end{equation}
where the scaling parameter $\epsilon_0$ is the well depth for the cross configuration
$(\hat{\bf e_i}.{\hat{\bf e_j}}=\hat\br_{ij}.\hat{\bf e_i}=\hat\br_{ij}.\hat{\bf e_j}=0)$.
The first of these functions 
\be
\epsilon ({\hat{\bf e_i}}, {\hat{\bf e_j}}) = [1 - \chi^2
({\hat{\bf e_i}}.{\hat{\bf e_j}})^2]^{-\frac{1}{2}}
\ee
favors the parallel alignment of the particle and so aids liquid crystal 
formations. The second function has a form analogous to 
$\sigma({\hat{\bf e_i}}, {\hat{\bf e_j}}, {\hat\br_{ij}})$, i.e.
\begin{widetext}
\ba
\epsilon'({\hat {\bf e_i}}, {\hat {\bf e_j}}, {\hat\br_{ij}}) =
\left[ 1 - \chi'\left(\frac{({\hat{\bf e_i}}.{\hat\br_{ij}})^2
+ ({\hat{\bf e_j}}.{\hat\br_{ij}})^2 - 2\chi'({\hat{\bf e_i}}.
{\hat\br_{ij}})({\hat{\bf e_j}}.{\hat\br_{ij}})
({\hat{\bf e_i}}.{\hat{\bf e_j}})}
{1 - \chi^{\prime 2}({\hat{\bf e_i}}.{\hat{\bf e_j}})^2}\right)
\right]
\ea
\end{widetext}
The parameter $\chi'$ is determined by the ratio of the well depth as 
\be
\chi' = \frac{k'^{ 1/\mu}-1}{k'^{ 1/\mu}+1}. 
\ee
Here $k'$ is well-depth ratio for the side-by-side and end-to-end configuration.

The GB model contains four parameters $(x_0, k', \mu, \nu)$ that determine 
the anisotropy in the repulsive and attractive forces in addition to two 
parameters $(\sigma_0, \epsilon_0)$ that scale the distance and energy, respectively.
Though $x_0$ measures the anisotropy of the repulsive core, it also determines the 
difference in the depth of the attractive well between the side-by-side and the cross
configurations. Both $x_0$ and $k'$ play important role in stabilizing the liquid crystalline phases.
The exact role of other two parameters $\mu$ and $\nu$ are not very obvious; though 
they appear to affect the anisotropic attractive forces in 
a subtle way. 

The phase diagram found for the system interacting via GB potential of Eqs. (1.1)-(1.8)
exhibits isotropic, nematic and Sm $B$ phases \cite{4,5} for $x_0=3.0, k'=5.0, \mu=2$, and $\nu=1$.
An island of Sm $A$ is, however, found to appear in the  phase diagram  
at value of $x_0$ slightly greater than 3.0 \cite{5}. The range of Sm $A$ 
extends to both higher and lower temperatures
as $x_0$ is increased. Also as $x_0$ is increased, the isotropic-nematic ($I-N$) 
transition is seen to move to lower density (and pressure) at a given temperature. 
Bates and Luckhurst \cite{6} have investigated the phases and phase transition for the 
GB potential with $x_0=4.4, k'=20.0, \mu=1$ and $\nu=1$ using the isothermal-isobaric Monte-Carlo
simulations. At low pressure they found isotropic, Sm $A$ and Sm $B$ phases but not the 
nematic phase. However as the pressure is increased, the nematic phase also get 
stabilized and a sequence of $I-N$-Sm $A$ and Sm $B$ was found.

In this paper we use the density functional approach to examine the different phases 
using Gay-Berne potential with the parameters chosen in the ref. \cite{6} and compare 
our results with the results found by the simulations. The paper is organized as follows.
In Sec. II, we describe Percus-Yevick integral equation theory for the calculation of 
the pair correlation functions of the isotropic phase. We compare our results with
those found by simulations. In Sec. III the density-functional formalism has been 
used to locate the freezing transitions and freezing parameters. The paper ends with
a discussion given in Sec. IV.  
\section{\bf Isotropic liquid: Pair Correlation Functions} 
In the theory of freezing to be discussed in the next section the structural
informations of isotropic phase will be used as input data.
The structural information of an isotropic liquid is contained in the two particle 
density distribution $\rho(\bn,\bp)$ as the single particle density distribution is constant
independent of position and orientation. The two-particle 
density distribution $\rho(\bn,\bp)$ measures the probability of finding simultaneously
a molecule in a volume element d$\br_1$d$\bom_1$ centered at$(\br_1,\bom_1)$ 
and a second molecule in a volume element d$\br_2$d$\bom_2$ at $(\br_2,\bom_2)$. The pair 
correlation function $g(\bn,\bp)$ is related to $\rho(\bn,\bp)$ as 
\be
g(\bn,\bp)=\frac{\rho(\bn,\bp)}{{\rho(\bn)}{\rho(\bp)}}
\ee 
where $\rho(\bf i)$ is the single particle density distribution. Since for 
the isotropic fluid $\rho(\bn)=\rho(\bp)=\rho_f=\langle N \rangle/V$, where $\langle N \rangle$
is the average number of molecules in the volume V,
\be
\rho_{f}^{2}g(\br, \bom_1, \bom_2)= \rho(\br, \bom_1, \bom_2)
\ee
where $\br=\br_2-\br_1$. In the isotropic phase $\rho({\bn,\bp})$ depends only 
on distance $|\br_2-\br_1|=r$, the orientation of molecules with respect to each other
and on the direction of vector $\br$. 

The pair distribution function $g(\bn,\bp)$ of the isotropic fluid is of particular 
interest as it is the lowest order microscopic quantity that contains information about
the translational and orientational structures of the system and also has direct contact 
with intermolecular (as well as with intramolecular) interactions. For an ordered phase,
on the other hand, as shown in the next section most of the 
structural informations are contained in the single 
particle distribution $\rho({\bn})$. In the density functional theory of freezing 
the single particle distribution $\rho(\bn)$ of an ordered phase is expressed in terms of the 
pair correlation function of the isotropic fluid (see Sec III). 

The value of $g(\bn,\bp)$ as a function of intermolecular separation and orientation
at a given temperature and density is found  either by computer simulation or by
solving the Ornstein-Zernike(OZ) equation
\begin{eqnarray}
h(\bn, \bp) = c(\bn, \bp)+\rho_f\int c(\bn, \bs) h(\bp, \bs) d\bs
\end{eqnarray}
where $d\bs = d\br_3 d\bom_3$ and $h(\bn, \bp) = g(\bn, \bp) - 1$ and
$c(\bn, \bp)$ are, respectively, the total and direct pair correlation
functions(DCF), using a suitable closure relation such as \PY (PY) integral equation
, hypernetted chain (HNC) relations. Approximations are introduced through
these closure relations \cite{7}.

The Percus-Yevick closure relation is written in various equivalent form.The
form adopted here is
\be\
c^{PY}(\bn, \bp) = f(\bn, \bp)[g(\bn, \bp)-c(\bn, \bp)]
\ee\
where $f(\bn, \bp)=\exp[-\beta u(\bn, \bp)]-1$ is Mayer function , $\beta=(k_BT)^{-1}$ and
$u(\bn, \bp)$ is a pair potential of interaction. Since for the isotropic liquid 
DCF is an invariant pair wise function, it has an expansion in body fixed (BF) 
frame in terms of basic set of rotational invariants, as
\be\
c(r_{12},\bom_1,\bom_2)=\sum_{l_{1}l_{2}m}c_{l_1l_{2}m}(r_{12})Y_{l_{1}m}(\bom_1)
Y_{l_{2}{\underline m}}(\bom_2)
\ee\
where ${\underline m}$=-m. The coefficients $c_{l_1l_{2}m}(r_{12})$ are defined as
\be\
c_{l_1l_{2}m}(r_{12})=\int c(r_{12},\bom_1,\bom_2) Y^*_{l_1m}(\bom_1)
Y_{l_2{\underline m}}^{*}(\bom_2)d\bom_1d\bom_2
\ee\
Expanding all the angle dependent functions in BF frame, the OZ equation reduces to a set of
algebraic equation in Fourier space
\be\
h_{l_1l_{2}m}(k)=c_{l_1l_{2}m}(k)+(-1)^{m}\frac{\rho_f}{4\pi}\sum_{l_3}
c_{l_1l_{3}m}(k)h_{l_3l_{2}m}(k)
\ee\
where the summation is over allowed values of $l_3$. The PY closure relation is expanded
in spherical harmonics in the body (or space) fixed frame. The pair correlation functions are then
found by solving these self consistently \cite{8}.

In our earlier work \cite{9,10} we considered $30$ harmonics in expansion of each orientation dependent 
function (see Eq.(2.5)); i.e. the series were truncated at the value
of $l$ indices equal to $6$. Since the accuracy of the results depends on this 
number and as the anisotropy of the shape taken here is larger than the earlier 
work, we considered $54$ harmonics. The series of each orientation 
dependent function was truncated at the value of $l$ indices equal to 8. The numerical procedure
for solving Eq(2.7) is the same as discussed in ref.\cite{10}. 

In Fig.1 we compare the values of g(r*)=1+$h_{000}(r*)/4\pi$ in BF-frame  having 30 and
54 harmonic coefficients at $T^*(=k_BT/\epsilon_0)=1.40$ and density 
$\eta(={\pi}\rho_{f}\sigma_0^3x_0/6)$= 0.44
for $x_0=4.4$, where $r^*=r/\sigma_0$ is the reduced interparticle separation. 
One other projection of the pair-correlation 
is shown in Fig.2 for the same set of parameters. It is obvious from these figures 
that even for $x_0=4.4$ one gets good results with 30 harmonics.  

In Figs.3 and 4 we compare the values of $g(r^*)$ and $g_{220}(r^*)$, respectively,
with those obtained by computer simulations \cite{6} for $\eta=0.36$ and $T^*=1.80$.
From these figures we find that while the PY theory gives qualitatively 
correct results for the pair correlation functions, quantitatively it underestimates 
the orientational correlations. The PY peak in $g(r^*)$ (see Fig.3) is broad and 
of less hight than the one found from the simulation. Though peak hight 
in $g_{220}(r^*)$ compare well, the oscillation at large $r^*$ is damped more in 
PY results than in the simulation data. However, as emphasized in our 
earlier work the PY theory is reasonably accurate for the GB potential
at low temperatures \cite{10}. 
\section{Freezing transitions}
In the density functional approach one uses the grand thermodynamic potential defined as
\begin{equation}
-W = \beta A - \beta {\mu_c} \int d{\bf x} \rho({\bf x})
\end{equation}
where A is the Helmholtz free energy, $\mu_c$ the chemical potential and
$\rho({\bf x})$ is a singlet distribution function, to locate the transition. It is convenient 
to subtract the isotropic fluid thermodynamic potential from $W$ and write it as \cite{11}
\begin{equation}
\Delta W = W - W_f = \Delta W_1 + \Delta W_2
\end{equation}
with
\begin{widetext}
\begin{eqnarray}
\frac{\Delta W_1}{N} & = & \frac{1}{\rho_f V} \int {d\bf r}
{d\bf \Omega}\left\{{\rho({\bf r}, {\bf \Omega})
\ln \left[\frac{\rho({\bf r}, {\bf \Omega})}{\rho_f}\right] - \Delta
\rho({\bf r}, {\bf \Omega})}\right\} \\
{\rm and} \nonumber \\
\frac{\Delta W_2}{N} & = & -\frac{1}{2\rho_f} \int {d\bf r_{12}}
{d\bf \Omega_1}{d\bf \Omega_2}\Delta\rho({\bf r_1}, {\bf \Omega_1})
c({\bf r_{12}}, {\bf \Omega_1}, {\bf \Omega_2})\Delta\rho({\bf r_2},
{\bf \Omega_2})
\end{eqnarray}
\end{widetext}
Here $\Delta\rho({\bf x}) = \rho({\bf x}) - \rho_f$, where
$\rho_f$ is the density of the coexisting liquid.

The minimization of $\Delta W$ with respect to arbitrary variation in the ordered phase density
subject to a constraint that corresponds to some specific feature of the ordered phase 
leads to  
\ba
\ln\frac{\rho({\bf r_1},{\bf \Omega_1})}{\rho_f} &=& \lambda_L +
\int d{\bf r_2} d{\bf \Omega_2} c({\bf r_{12}}, {\bf \Omega_1},
{\bf \Omega_2};\rho_f)\times\nonumber\\
&&\Delta\rho({\bf r_2}, {\bf \Omega_2})
\ea
where $\lambda_L$ is Lagrange multiplier which appears in the
equation because of constraint imposed on the minimization.

Eq.(3.5) is solved by expanding the singlet distribution $\rho({\bf x})$ in 
terms of the order parameters that characterize the ordered structures. One can use 
the Fourier series and Wigner rotation matrices to expand $\rho(\br,\bom)$. Thus
\be\
\rho(\br, \bom) = \rho_0 \sum_q \sum_{lmn} Q_{lmn}
(G_q) \exp(i{\bf G}_q.\br) D^l_{mn}(\bom)
\ee\
where the expansion coefficients
\ba\
 Q_{lmn}(G_q) &=& \frac{2l+1}{N}\int d\br \int d\bom \rho(\br, \bom)\times\nonumber\\
&&\exp(-i{\bf G_q}.\br) D^{*l}_{mn}(\bom)
\ea\
are the order parameters, ${\bf G_q}$ the reciprocal lattice vectors, $\rho_0$
the mean number density of the ordered phase and $D^{*l}_{mn}(\bom)$ the generalized spherical
harmonics or Wigner rotation matrices. Note that for a uniaxial system consisting 
of cylindrically symmetric molecules $m=n=0$ and, therefore, one has 
\be\
\rho(\br, \bom) = \rho_0 \sum_l \sum_q Q_{lq} \exp(i{\bf G}_q.\br)
                  P_l(\cos \theta)
\ee\
and
\be\
Q_{lq} = \frac{2l+1}{N}\int d\br \int d\bom \rho(\br, \bom)
\exp(-i{\bf G_q}.\br) P_l(\cos \theta)
\ee\
where $ P_l(\cos \theta)$ is the Legendre polynomial of degree $l$ and
$\theta$ is the angle between the cylindrical axis of a molecule and the
director.

In the present calculation we consider two orientational order parameters
\ba
\bar P_l&=&\frac{Q_{l0}}{2l+1}=\langle P_l(cos\theta)\rangle
\ea
with $l$=2 and 4, one order parameter corresponding to positional order 
along Z axis,
\ba
{\bar\mu}&=&Q_{00}(G_z)=\langle cos(\frac{2\pi}{d}z)\rangle
\ea
($d$, being the layer spacing) and one mixed order parameter that measures the 
coupling between the positional and orientational ordering and is defined as,
\ba
\tau&=&\frac{1}{5}Q_{20}(G_z)=\langle cos(\frac{2\pi}{d}z) P_{l}(cos\theta)\rangle
\ea
The angular bracket in above equations indicate the ensemble average.

The following order parameter equations are obtained by using Eqs.(3.5) and (3.9):
\begin{widetext}
\ba
\bar P_l &=&\frac{1}{2d}\int_{0}^{d}dz_{1}\int_{0}^{\pi}sin\theta_1 d\theta_1
P_l(cos\theta_1)\exp[sum]\\
{\bar\mu} &=&  
 \frac{1}{2d}\int_{0}^{d}dz_{1}\int_{0}^{\pi}sin\theta_1 d\theta_1 
cos(\frac {2\pi z_1}{d})\exp[sum]\\
\tau &=&
\frac{1}{2d}\int_{0}^{d}dz_{1}\int_{0}^{\pi}sin\theta_1 d\theta_1
P_2(cos\theta_1)cos(\frac{2\pi z_1}{d})
\exp[sum]
\ea
and change in density at the transition is found from the relation 
\be
1+\Delta\rho^{*}=\frac{1}{2d}\int_{0}^{d}dz_{1}\int_{0}^{\pi}sin\theta_1 d\theta_1
\exp[sum].
\ee
Here 
\ba
sum &=& \Delta\rho^{*}\hat C_{00}^0+2{\bar\mu} cos(\frac{2\pi z}{d})\hat C_{00}^1(\theta_1)+
\bar P_2 \hat C_{20}^{0}(\theta_1)+ 
\bar P_4 \hat C_{40}^{0}(\theta_1)+\nonumber\\
&&2\tau cos(\frac{2\pi z}{d}) \hat C_{20}^{1}(\theta_1)
\ea
and 
\ba
\hat C_{L0}^{q}(\theta_1)&=&({\frac{2l+1}{4\pi}})^{1/2}\rho_f\sum_{l_1l}i^l(2l_1+1)^{1/2}(2l+1)^{1/2}
C_g(l_1Ll;000)P_{l_{1}}(cos\theta_1)\times\nonumber\\
&&\int_{0}^{\infty}c_{l_1Ll}(r_{12}) j_{l}(G_{q}r_{12})r_{12}^2 dr_{12}
\ea
\end{widetext}
where $C_{g}(l_1Ll;000)$ are Clebsch-Gordon coefficients and $G_q=2\pi/d$. 

In the isotropic phase all the four order parameters become zero. In the nematic phase the
orientational order parameters $\bar P_2$ and $\bar P_4$ become nonzero but the other two
parameters $\bar\mu$ and $\tau$ remain zero. This is because the nematic phase has no long range 
positional order. In the Sm $A$ phase all the four order parameters are nonzero showing that 
the system has both the long range orientational and positional order along 
one direction. Equations (3.10)-(3.16) are solved self consistently using the values of direct
pair correlation function harmonics $c_{l_{1}l_{2}l}(r)$ evaluated in the previous section
at a given value of T* and $\eta$. This calculation is repeated with different values of $d$,
the interlayer spacing. By substituting these solutions in Eqs.(3.2)-(3.4) we find the
grand thermodynamic potential difference between ordered and isotropic phases; i.e.
\begin{widetext}
\ba
-\frac{\Delta W}{N}&=& -\Delta\rho^{*}+\frac{1}{2}\Delta\rho^{*}(2+\Delta\rho^{*})\hat C_{00}^{0}+
\frac{1}{2}(\bar P_{2}^{2}\hat C_{22}^{0}+\bar P_{4}^{2}\hat C_{44}^{0})+
 {\bar\mu}^{2}\hat C_{00}^{1}+\nonumber\\
&&2{\bar\mu}\tau\hat C_{20}^{1}+\tau^{2}\hat C_{22}^{1}
\ea
where
\ba
\hat C_{LL'}^{q}&=&(2L+1)^{1/2}(2L'+1)^{1/2}\rho_f\sum_l i^l({\frac{2l+1}{4\pi}})^{1/2}
C_g(LL'l;000)\times\nonumber\\
&&\int_{0}^{\infty} c_{LL'l}(r_{12}) j_{l}({ G_q r_{12}}) r_{12}^{2} dr_{12}
\ea
\end{widetext}
At a given temperature and density a phase with lowest grand potential is taken as 
the stable phase. Phase coexistence occurs at the values of $\rho_f$ that makes 
${-\Delta W}/{N}=0$ for the ordered and the liquid phases. The transition
from nematic to the Sm $A$ is determined by comparing the values of $-{\Delta W}/{N}$
of these two phases at a given temperature and at different densities. We have 
done this analysis at four different temperatures, i.e. at $T^*$=1.2, 1.4, 1.6 and 1.8.
Our results are summarized in Table 1. We see that at low temperatures, i.e. at 
$T^*$=1.2 and 1.4 the isotropic liquid freezes directly into Sm $A$ on increasing the density.
Nematic phase is not stable at these temperatures. However, at $T^*$=1.6 the isotropic liquid 
on increasing the density is found first to freeze into the nematic phase at 
$\eta=0.49$ and on further increasing the density the nematic phase transforms into
Sm $A$ phase at $\eta= 0.519$. At high temperature, $T^*$=1.8 the transition is found to take place
between isotropic and nematic only for the density range considered by us.
\begin{table*}
\caption{Values of the order parameters at the transitions for the GB potential
with $x_0=4.4, k'=20, \mu=\nu=1$.
Quantities in  reduced units are $d^*=d/\sigma_0$,
 Pressure $P^* = P\sigma_0^3/\epsilon_0,
\mu^{*}_c = \mu_{c}/\epsilon_0,
\;\; {\rm and} \;\; \eta = \pi\rho_f\sigma_0^3 x_0/6$}
\label{tab2}
\begin{ruledtabular}
\begin{tabular}{ccccccccccc}
\small
 $T^*$&Transition&$\eta$ & $d^*$ & $\Delta \rho^*$ &$\bar\mu $ &  ${\bar P_2}$ & ${\bar P_4}$ &
 $\tau$ & $P^*$ & $\mu^{*}_{c}$ \\  \hline
1.20&I-Sm A &0.390  &3.67&0.212 &0.64 &0.99 &0.71&0.52  &1.06&6.51 \\
1.40&I-Sm A &0.440  &3.90 &0.164 &0.65 &0.95 &0.66 &0.54&1.72&10.54 \\
1.60&I-N & 0.490&0.0 & 0.087&0.0 & 0.76 & 0.49 &0.0& 2.61 & 15.49 \\
&N-Sm A & 0.519&4.00 &0.127 &0.49&0.94 &0.67 &0.42 &3.01&17.34 \\
1.80 &I-N & 0.530&0.0 & 0.072&0.0 & 0.75 & 0.47&0.0 & 3.57 &20.53
\end{tabular}
\end{ruledtabular}
\end{table*}                 

Using the results summarized in Table 1 we draw the phase digram in Fig.5 in the 
density-temperature plane. The full circles represent the points given in Table 1 and
lines are drawn to demarcate the phase boundaries. 
\section{Discussions}
The phase diagram shown in Fig.5 is in good qualitative agreement 
with the one found by computer simulations \cite{6}. At low temperature the 
nematic is unstable; on increasing the density the fluid freezes directly 
into Sm $A$ phase. The nematic phase is found to stabilize in-between the 
isotropic and Sm $A$ phases when $T^{*}\ge$1.5.
When we compare the phase diagram given in Fig.5 with the one reported by 
Bates and Luckhurst \cite{6} we find that while $N$-Sm $A$ boundary is in good
agreement, the $I-N$ boundary is shifted towards lower temperatures 
compared to that given in ref.\cite{6}. As a consequence the region of existence 
of the nematic phase in the phase diagram found by us is relatively narrow. 
For example, at $P^*$=3.0 we find $T_{N-Sm A}/T_{I-N}=0.945$ whereas the value  
found by Bates and Luckhurst \cite{6} is 0.795.
We also find that the values of the orientational order parameters and the 
change in density at the transitions are higher than those given in ref \cite{6}.

The reason for finding the $I-N$ phase boundary at lower temperatures 
compared to that found in simulation can be traced to the inadequacy of the 
PY integral equation theory discussed in Sec.II to estimate the values of
orientational correlations at higher temperatures. Moreover, as we have shown in our
earlier work \cite{10} that the second-order density functional method (discussed in
Sec.III) has the tendency to overestimate the orientational order parameters
and the change in density. The modified weighted density approximation (MWDA) \cite{10, 12}
version of density-functional method is expected to improve the agreement.
We have checked it by using this theory to locate the $I-N$ transition at 
T*=1.6 and 1.8. We find that $I-N$ transition at $T^*$=1.6 takes place at
$\eta=0.471$ with $\Delta\rho^*=0.042, \bar P_2=0.59, \bar P_4=0.35$ and
$P^*$=2.38. These values are in better agreement with computer simulation
than those given in Table 1. Similarly at $T^*$=1.8 the $I-N$ transition takes
place at $\eta=0.498$ with $\Delta\rho^*=0.038, \bar P_2=0.61, \bar P_4=0.35$ and
$P^*$=3.10 giving better agreement compared to the values found from
the second-order density-functional theory. The dashed line in Fig.5 
demarcating the $I-N$ boundary represent the results found from the 
MWDA version of the density functional method. If we take the dashed 
line as the phase boundary between isotropic and nematic phases then at 
$P^*$=3.0, $T_{N-Sm A}/T_{I-N}=0.890$, instead of 0.945.  

In the conclusions we wish to emphasize that the freezing transitions in
complex fluids can be predicted reasonably well with the density-functional 
method if the values of pair correlation functions in the isotropic phase 
are accurately known.
\section{acknowledgements}
The work was supported by the Department of Science
and Technology (India) through a project grant. 

\newpage
\begin{figure}[h]
\includegraphics[width=3.6in]{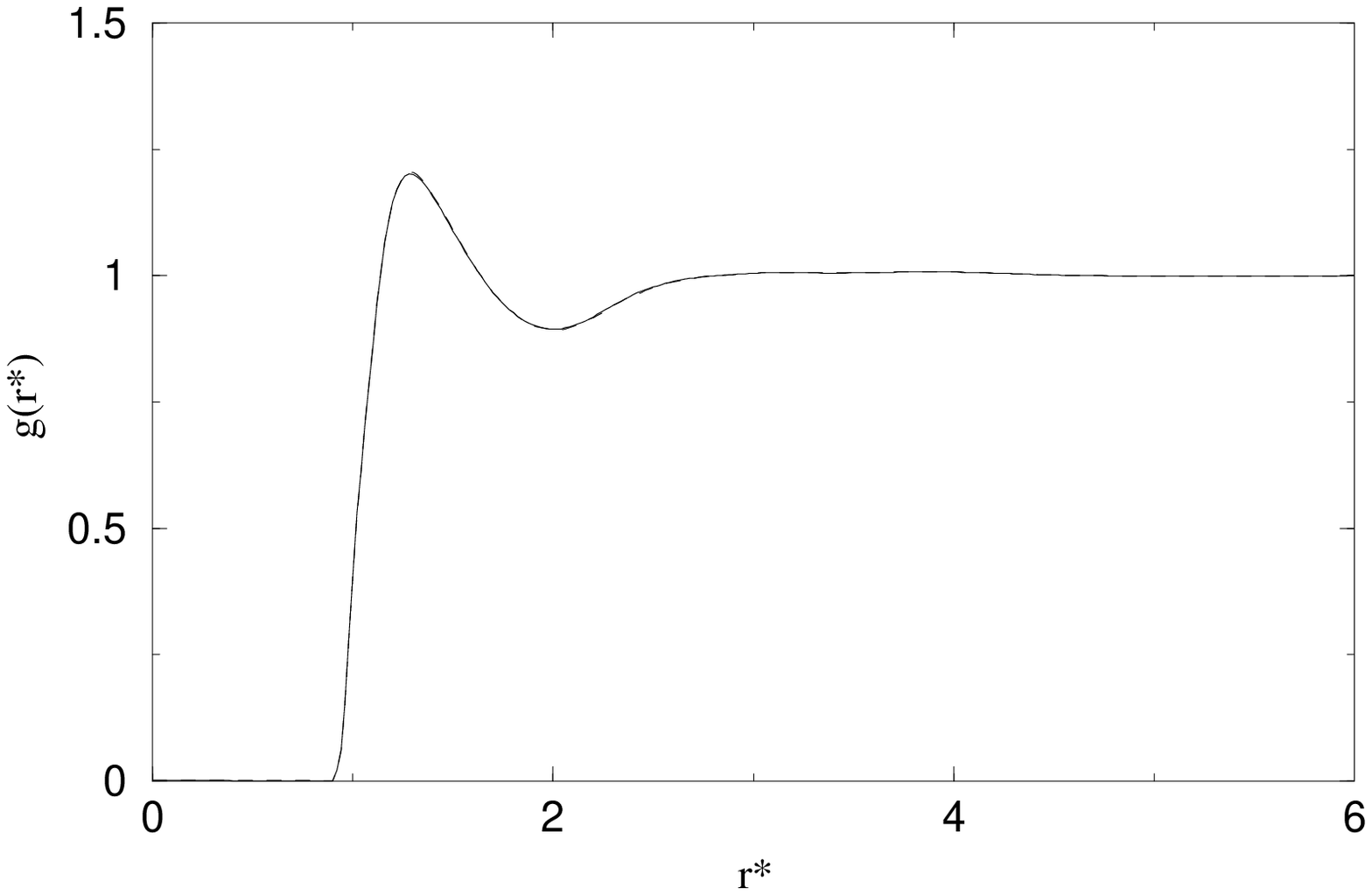}
\caption{Pair-Correlation functions of the centre of mass g($r^*$) for GB fluid
with parameters $x_0= 4.4, k'= 20.0, \mu=1$ and $\nu$ = 1, at  $\eta= 0.44$ and $T^*=1.40$.
The solid and dashed curves are, respectively for, 30 and 54 body-fixed harmonic
coefficients. These two curves are indistinguishable on the scale of the figure.}           
\end{figure}
\begin{figure}[h]
\includegraphics[width=3.6in]{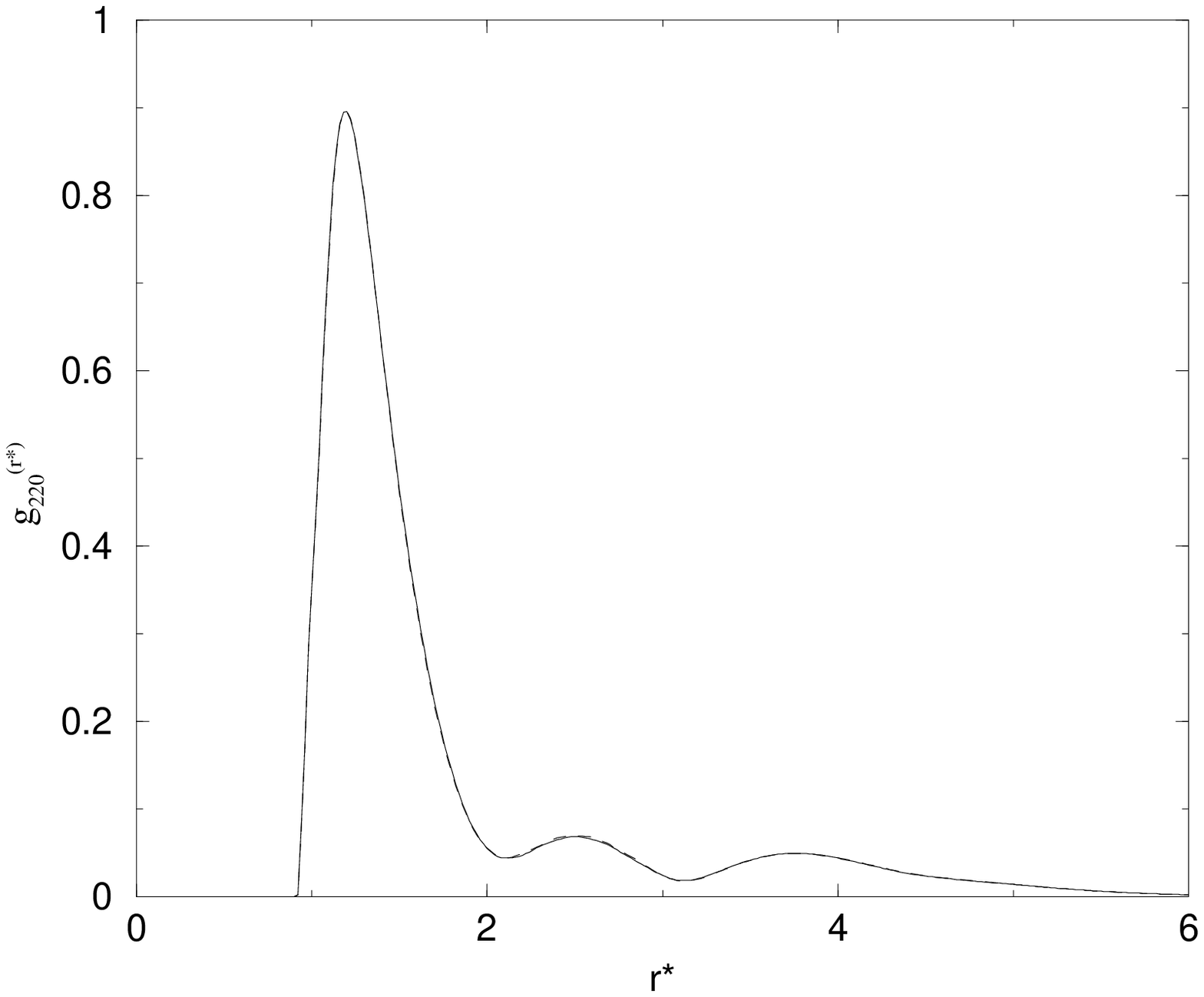}
\caption{Spherical harmonic coefficient $g_{220}(r^*)$ in the body-fixed frame.
The curves are the same as in Fig.1.}
\end{figure}                                                 
\begin{figure}[h]
\includegraphics[width=3.6in]{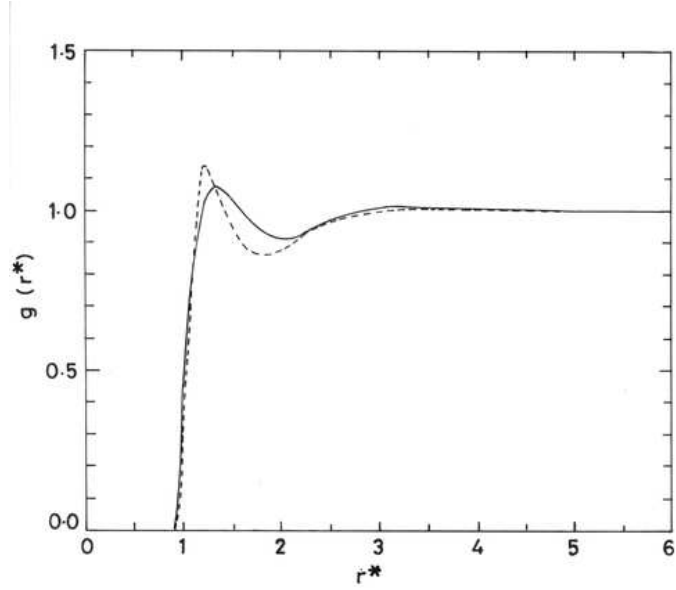}
\caption{Pair-Correlation function of the centre of mass g($r^*$) for GB fluid
with parameters $x_0= 4.4, k'= 20.0, \mu=1$ and $\nu$ = 1 at $\eta= 0.36$ and $T^*= 1.80$. The solid curve 
is our PY result and dashed curve is the simulation result of Bates and Luckhurst [6].}
\end{figure}                                                 
\begin{figure}[h]
\includegraphics[width=3.6in]{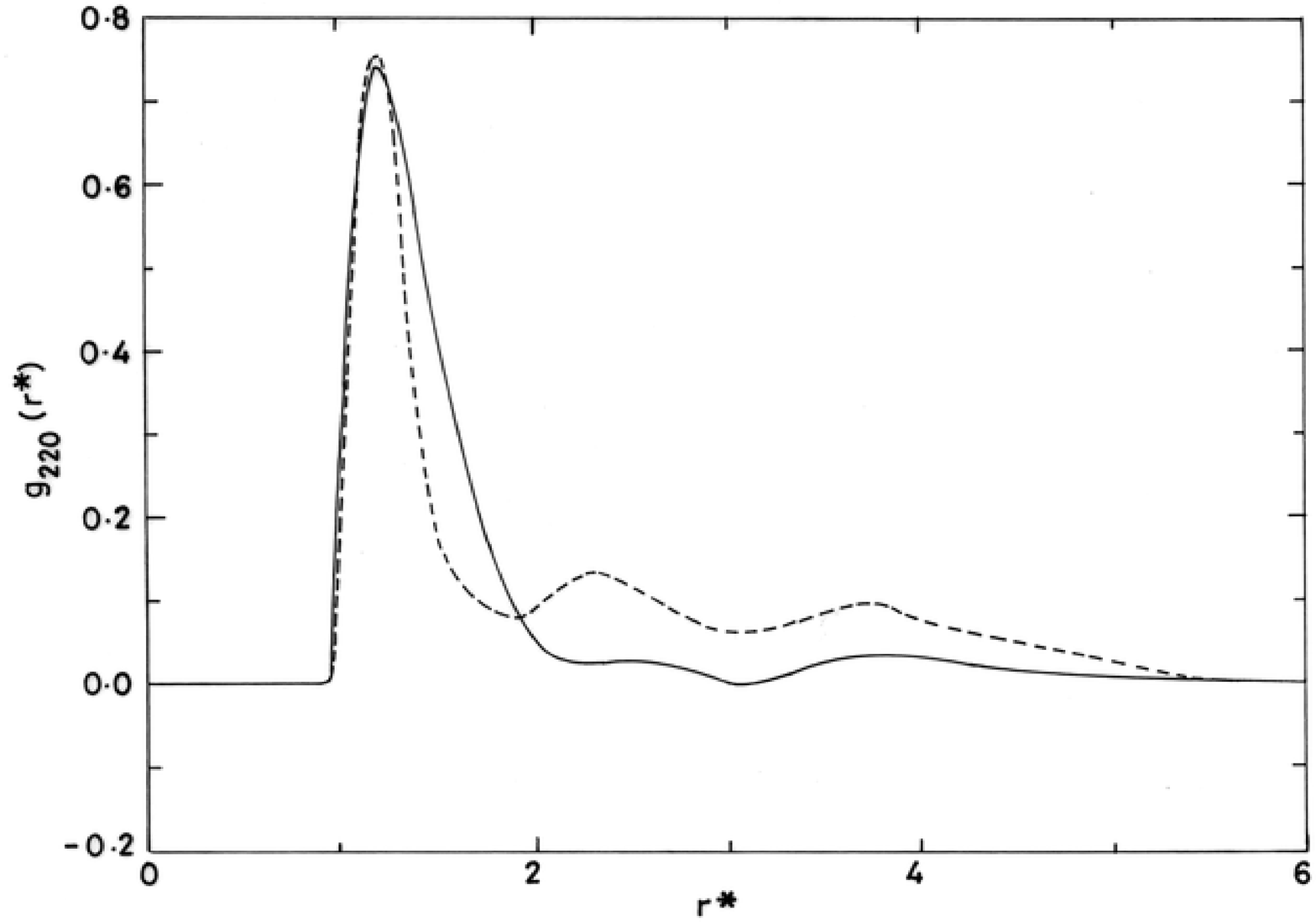}
\caption{Spherical harmonic coefficient $g_{220}(r^*)$ in the body-fixed frame. The 
curves are same as in Fig.3.}
\end{figure}                                                 
\begin{figure}[h]
\includegraphics[width=3.6in]{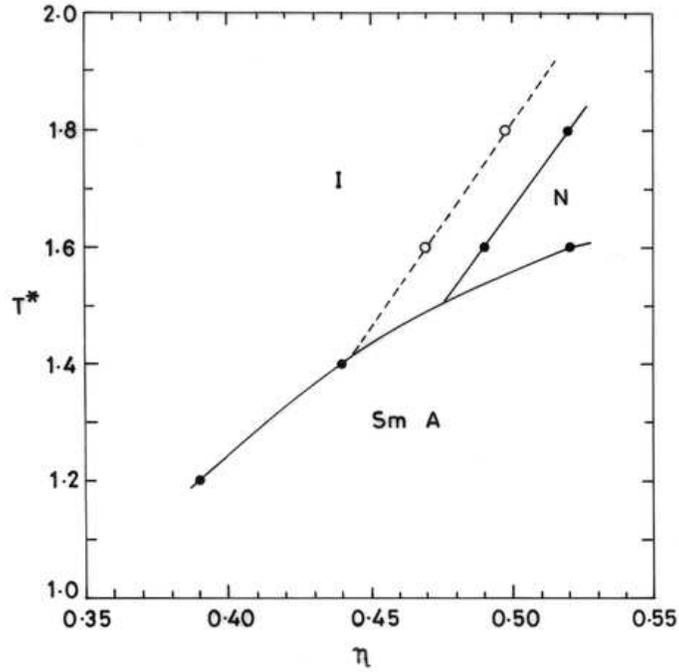}
\caption{Phase diagram for the GB potential with parameters $x_0= 4.4, k'= 20.0, \mu=1$
and $\nu$ = 1. The solid lines indicate the phase boundaries obtained by using
the second-order density-functional theory. The dashed line demarcating the isotropic
and nematic phase is found from the modified weighted density approximation version
of the density-functional theory.}       
\end{figure}                                                 
\end{document}